\documentclass[sigconf]{acmart}
\AtBeginDocument{%
  }

\usepackage{pifont}
\usepackage{float}
\usepackage{tabularx}
\usepackage{enumitem}
\usepackage{tcolorbox}
\usepackage{url}
\tcbuselibrary{skins}
\newcommand{\cmark}{\ding{51}}
\newcommand{\xmark}{\ding{55}}

\setcopyright{none}
\renewcommand\footnotetextcopyrightpermission[1]{}
\begin{document}

\title[Mono2Sls]{Mono2Sls: Automated Monolith-to-Serverless Migration via Multi-Stage Pipeline with Static Analysis}

\author{Xingyan Chen}
\email{chenxy979@mail2.sysu.edu.cn}
\affiliation{%
  \institution{Sun Yat-sen University}
  \city{Zhuhai}
  \country{China}
}

\author{Yuxin Su}
\email{suyx35@mail.sysu.edu.cn}
\affiliation{%
  \institution{Sun Yat-sen University}
  \city{Zhuhai}
  \country{China}
}

\author{Zishan Su}
\email{zshsu@link.cuhk.edu.hk}
\affiliation{%
  \institution{The Chinese University of Hong Kong}
  \city{Hong Kong}
  \country{China}
}

\author{Yang Yu}
\email{yuy@mail.sysu.edu.cn}
\affiliation{%
  \institution{Sun Yat-sen University}
  \city{Zhuhai}
  \country{China}
}

\author{Zibin Zheng}
\email{zhzibin@mail.sysu.edu.cn}
\affiliation{%
  \institution{Sun Yat-sen University}
  \city{Zhuhai}
  \country{China}
}
\renewcommand{\shortauthors}{Author et al.}

\begin{abstract}
Cloud computing platforms offer elastic scaling, managed infrastructure, and pay-per-use pricing, but moving existing monolithic backends to them remains a difficult software engineering task.
In practice, the migration requires coordinated changes to program structure, source code, infrastructure configuration, and cloud-specific design decisions, and these changes are still largely carried out by hand.
In this paper, we present \textsc{Mono2Sls}, an automated pipeline that converts monolithic web backends into deployable AWS SAM applications.
The pipeline combines lightweight static analysis of entry points, call graphs, and asynchronous behavior with four sequential tool-using LLM agents: Architect, Code Developer, SAM Engineer, and Consistency Validator. These agents communicate through explicit intermediate artifacts and consult a curated SAM knowledge base.
Evaluated on six benchmark applications totaling more than 10K lines of code and 76 business endpoints, \textsc{Mono2Sls} achieves 100\% deployment success without manual fixes.
It also reaches 66.1\% end-to-end correctness and 98.7\% API-coverage F1, whereas the commercial baselines achieve 53.7--61.2\% and 88.4\%, respectively.
The migrated systems show more consistent use of AWS-native authentication and asynchronous patterns, and an ablation study indicates that static-analysis-guided architecture planning contributes 23.4 percentage points to end-to-end correctness.
\end{abstract}



\maketitle

\section{Introduction}
\label{sec:intro}

Serverless computing is now widely used for cloud applications because it offers automatic scaling, pay-per-use pricing, and managed infrastructure. Industry reports estimate that the global serverless market will grow from \$21.9 billion in 2024 to \$44.7 billion by 2029 at a 15.3\% compound annual growth rate~\cite{globenewswire2026}, with 72\% of enterprises already running serverless workloads in production~\cite{calyo2026serverless}. AWS Lambda is central to this ecosystem~\cite{aws-lambda}. At the same time, much production software still runs as monoliths, and legacy architectures remain a major source of enterprise technical debt~\cite{protiviti2023techdebt}. Turning these systems into serverless applications is difficult to automate. An automated migration method must recover decomposition boundaries from tightly coupled code, preserve API contracts and business logic, redesign data access for a distributed runtime, and generate cloud-specific Infrastructure-as-Code (IaC) that remains consistent with the transformed code. Prior studies report recurring problems in testing event-driven systems, integrating legacy infrastructure, and finding reliable migration practices, especially when teams lack serverless experience~\cite{hamza2023serverless,wen2021serverless}. These coupled requirements motivate methods that combine program analysis with structured generation.

LLMs have shown strong performance on code generation tasks, from function-level synthesis to repository-level modification~\cite{hou2024llmse,jiang2024codellm}. Benchmarks such as SWE-Bench~\cite{swebench2024} and RepoGenesis~\cite{repogenesis} suggest that commercial assistants such as Cursor and Claude Code can already make substantial repository-scale changes. This progress makes LLMs a natural starting point for migration automation, but existing assistants are not designed to maintain the long-range consistency that serverless migration requires. Decisions about function boundaries, API definitions, handler implementations, and infrastructure declarations must remain aligned across multiple generated artifacts, which is difficult to achieve in a single loosely structured generation context. The difficulty is compounded by rapidly evolving cloud APIs: code LLMs often misuse infrequent APIs~\cite{jain2024apidoc}, and this risk is particularly relevant to AWS SAM, whose configuration space spans more than 800 resource types~\cite{wen2024slsdetector}.

Our approach, \textsc{Mono2Sls}, is a multi-stage pipeline for migrating monolithic web backends to AWS SAM applications. It begins by extracting structural facts from the monolith with static analysis and then feeds them into four sequential stages, each handled by a dedicated tool-using LLM agent (Architect, Code Developer, SAM Engineer, Consistency Validator). Together, these stages plan the target architecture, generate Lambda handlers and SAM templates, and verify consistency across artifacts. Decomposing the task in this way lets the system coordinate decomposition, code transformation, and infrastructure generation without forcing all decisions into a single generation context.

Our contributions are as follows:
\begin{itemize}[noitemsep,topsep=2pt,leftmargin=*]
    \item \textbf{A multi-stage migration pipeline}: A four-stage workflow in which dedicated tool-using LLM agents (Architect, Code Developer, SAM Engineer, Consistency Validator) handle architecture planning, code generation, infrastructure generation, and validation through explicit intermediate artifacts and isolated contexts.
    \item \textbf{Static-analysis-driven architecture planning}: Lightweight analysis that extracts HTTP entry points, cross-file call graphs with async semantics, and DynamoDB schema information to ground migration decisions.
    \item \textbf{Domain-grounded SAM generation and validation}: A curated SAM knowledge base together with a Consistency Validator that performs 11 cross-artifact checks to improve template correctness and deployment readiness.
    \item \textbf{A benchmark and empirical evaluation}: Six web applications reverse-engineered from AWS reference serverless implementations, evaluated for deployability, functional correctness, and cloud-native design adoption.
\end{itemize}

\section{Background and Related Work}
\label{sec:background}

\subsection{AWS Serverless Architecture}

Recent reviews of cloud computing identify migration from monolithic systems as an open problem in serverless computing~\cite{wen2023serverless,hellerstein2019serverless}. Empirical studies further show that configuration management and event-driven design are common sources of errors in practice~\cite{eismann2021serverless,wen2021serverless}. AWS Serverless Application Model (SAM)~\cite{aws-sam-guide} is a widely used Infrastructure-as-Code framework for AWS serverless applications. It describes Lambda functions~\cite{aws-lambda}, API Gateway, DynamoDB/S3, Cognito~\cite{aws-cognito}, and SQS~\cite{aws-sqs}/EventBridge~\cite{aws-eventbridge} in a YAML template that extends CloudFormation. SAM is expressive, but it is also unforgiving: small mistakes, such as invalid CORS combinations, reserved environment variables, or missing IAM policies, can block deployment. Template correctness is therefore a key requirement for any automated migration system~\cite{wen2024slsdetector}.

\subsection{LLM-based Code Generation and Migration}

LLM-based code generation has progressed from function-level benchmarks (HumanEval~\cite{chen2021humaneval}, MBPP~\cite{austin2021mbpp}) to repository-level tasks (SWE-Bench~\cite{swebench2024}, RepoGenesis~\cite{repogenesis}). Static analysis often improves repository-scale performance: STALL+~\cite{stall2024} shows that file-level dependency augmentation outperforms RAG~\cite{lewis2020rag} alone, and CodeAgent \cite{codeagent2024} shows that tool-equipped agents outperform pure-LLM inference at repository scale. For code \emph{translation}, AlphaTrans~\cite{alphatrans2025} decomposes repository-level Java-to-Python translation into reverse-call-order fragments with multi-level validation, achieving 96.4\% syntactic correctness across ten projects. These systems address \emph{peer-language translation}, that is, functionally similar implementations in another language. They do not address \emph{architectural migration}, which requires 
simultaneously redesigning application structure, replacing the runtime model, and 
authoring cloud-native IaC declarations.

\subsection{Program Migration and Modernization}

\textbf{Microservice decomposition.} Wang \emph{et al.}~\cite{wang2024microservice} benchmark eight decomposition tools and find that structural and semantic metrics can yield quite different service boundaries; Romani \emph{et al.}~\cite{romani2022microservice} propose a data-centric identification process for legacy systems; Kalia \emph{et al.}~\cite{kalia2020mono2micro} use AI-driven runtime call graphs to partition Java monoliths automatically. All three lines of work reason about \emph{logical} boundaries but stop short of producing deployable infrastructure, so cloud migration remains a manual downstream step.

\textbf{FaaSification.} Podilizer~\cite{spillner2017podilizer} and Node2FaaS~\cite{carvalho2023node2faas} mechanically extract Java/Node.js methods into Lambda functions via static analysis and code rewriting, without reasoning about architecture, authentication patterns, or asynchronous communication. RADF~\cite{radf2023} adds business-logic-aware clustering but still produces blueprints and provisional structures rather than executable Lambda handlers and SAM templates.

\begin{figure*}[!t]
    \centering
    \includegraphics[width=0.9\textwidth]{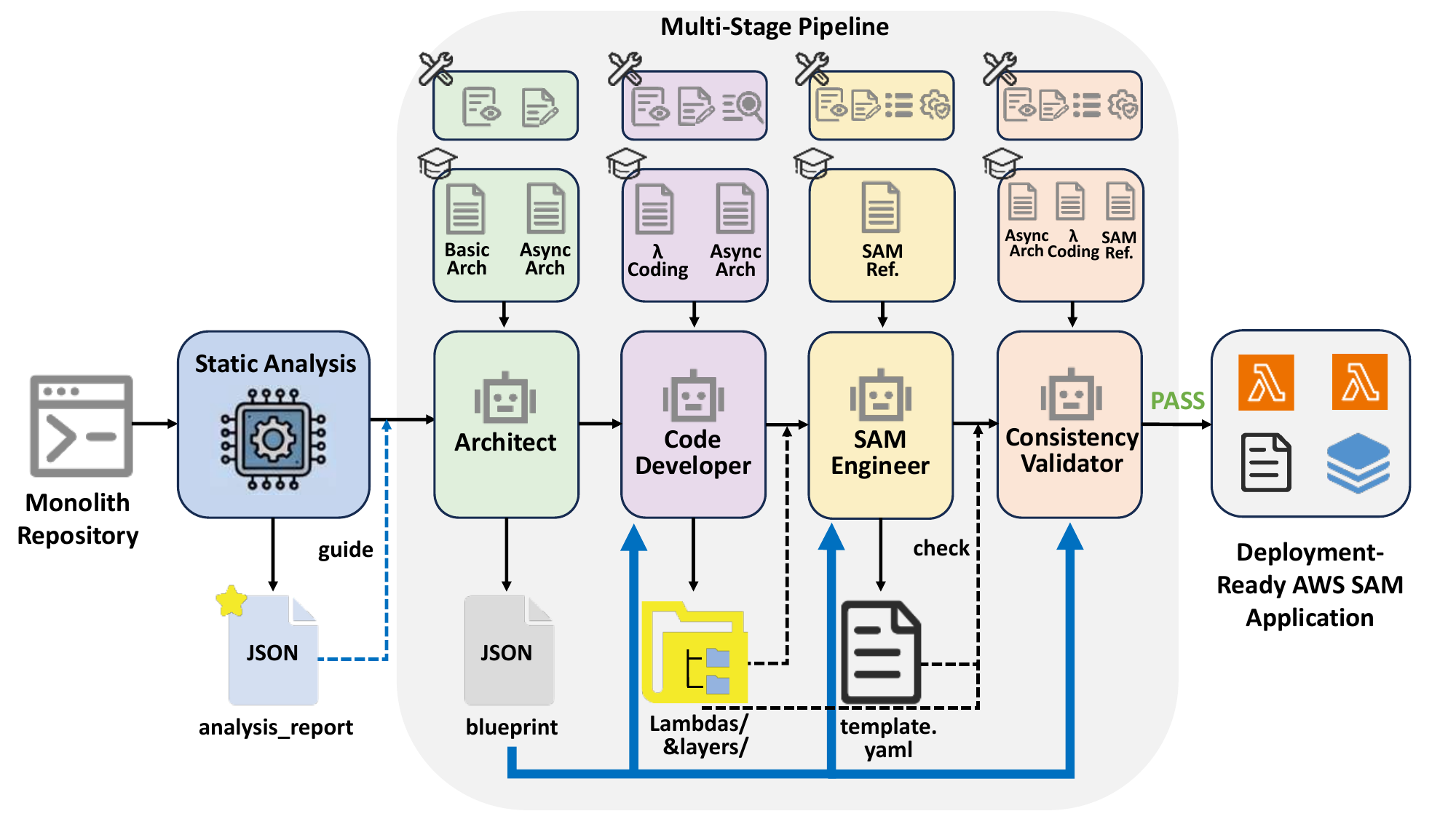}
    \caption{Mono2Sls pipeline overview. Static analysis preprocessing produces \texttt{analysis\_report.json}, which drives the Architect stage to design \texttt{blueprint.json}. The blueprint then serves as the shared contract for the three downstream stages, namely Code Developer, SAM Engineer, and Consistency Validator, each producing a well-defined output artefact.}
    \Description{The Mono2Sls pipeline: static analysis preprocessing feeds into four sequential stages, namely Architect, Code Developer, SAM Engineer, and Consistency Validator, each producing a well-defined output artefact.}
    \label{fig:pipeline}
\end{figure*}

\section{Approach}
\label{sec:approach}

\subsection{Overview}
\textsc{Mono2Sls} is a \emph{static-analysis-driven} multi-stage pipeline that automatically migrates monolithic web applications to AWS SAM applications. As shown in Figure~\ref{fig:pipeline}, static analysis first produces \texttt{analysis\_report.json}, which anchors all downstream LLM reasoning. The pipeline then executes four sequential stages, each powered by a dedicated tool-using LLM agent: the \textbf{Architect} consumes this report and emits \texttt{blueprint.json}, the shared architectural contract for subsequent stages. The \textbf{Code Developer} and \textbf{SAM Engineer} generate Lambda handlers and \texttt{template.yaml} respectively, while the \textbf{Consistency Validator} cross-checks all artifacts, applies fixes, and re-validates. Each agent is equipped with purpose-built tools and a curated domain knowledge base (\S\ref{sec:tools-knowledge}); stages communicate exclusively through files, with no direct inter-agent messaging.

\subsection{Static Analysis}
\label{sec:static-analysis}

\begin{figure*}[!t]
    \centering
    \includegraphics[width=\textwidth]{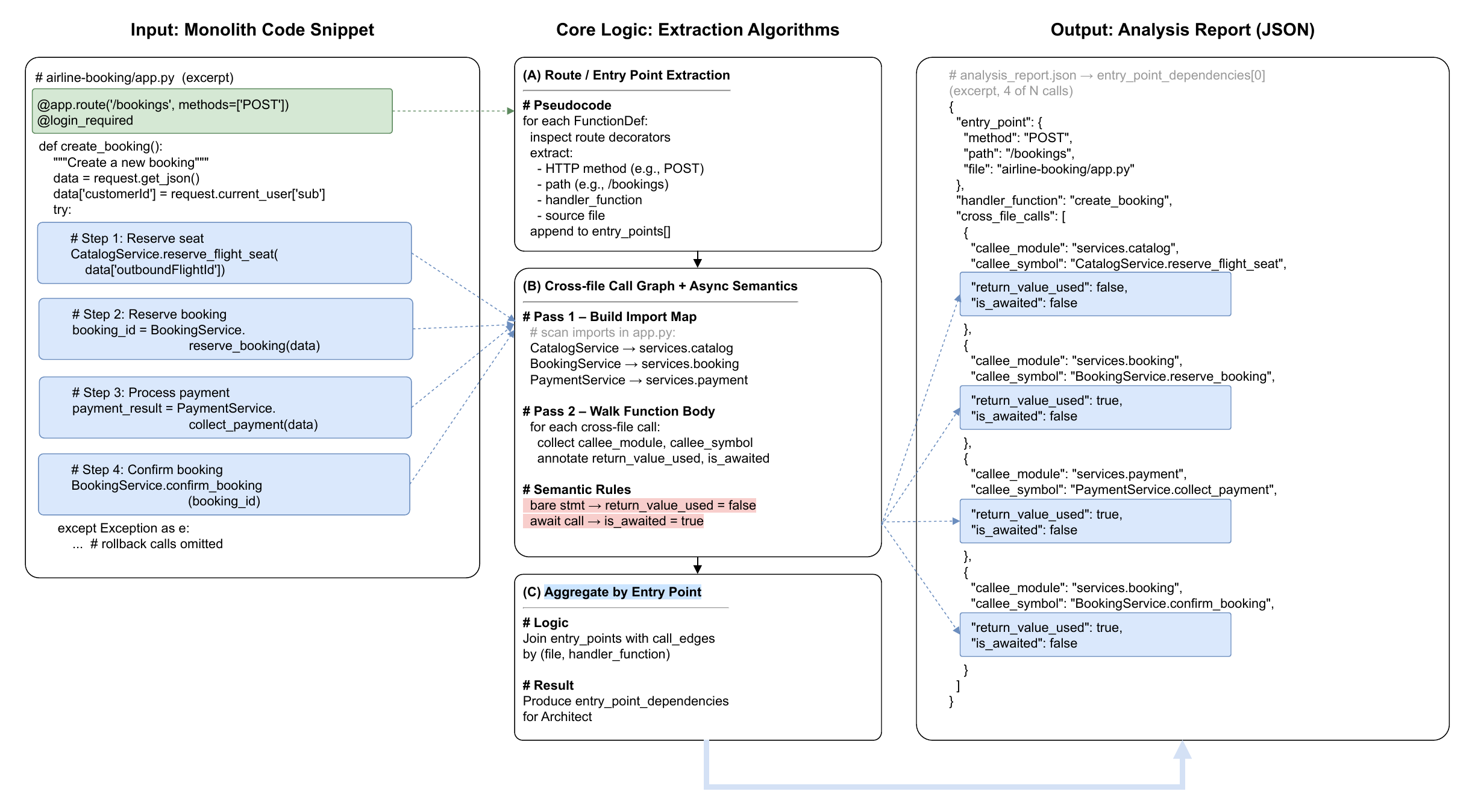}
    \caption{Static analysis and dependency extraction workflow. Left: monolith source snippet with annotated cross-file calls. Middle: core extraction algorithms for entry points, call graph, and async semantics. Right: the resulting \texttt{entry\_point\_dependencies} structure in \texttt{analysis\_report.json}.}
    \Description{Three-panel illustration of the static analysis process: left panel shows example Python Flask route handler code with a cross-file function call; middle panel shows AST-based extraction identifying the route decorator and the call edge annotated with return\_value\_used and is\_awaited flags; right panel shows the resulting JSON entry in entry\_point\_dependencies within analysis\_report.json.}
    \label{fig:static-analysis}
\end{figure*}

Before invoking the LLM-based agents, we perform lightweight static analysis to
extract structured information that anchors LLM reasoning. The analyzer produces
\texttt{analysis\_report.json}, which serves as the authoritative input for the
Architect agent, and a separate \texttt{symbol\_table.json} consumed exclusively
by \texttt{CodeRAGTool} for code indexing (\S\ref{sec:tools-knowledge}).

\textbf{HTTP Entry Points.} For Python, we traverse the AST to detect route
decorators (\texttt{@app.route}, \texttt{@app.get}, Flask Blueprint
prefix+path aggregation); for JavaScript/TypeScript, we apply regex
matching on Express patterns (\texttt{app.get}, \texttt{router.post}, etc.).
Each entry point records \texttt{(method, path, handler\_function, file)}.

\textbf{File Tags.} We classify each source file by content features, assigning
tags from \{\texttt{AWS\_SDK}, \texttt{DynamoDB}, \texttt{Auth}, \texttt{FileUpload}\}.
Tags are detected via keyword matching on the file source (e.g., \texttt{boto3}
imports for \texttt{AWS\_SDK}, DynamoDB API calls for \texttt{DynamoDB}, JWT
references for \texttt{Auth}). These tags guide the DynamoDB Schema Locator and
are available for downstream agent decisions.

\textbf{Cross-file Call Graph with Async Semantics.} A two-pass analysis builds a per-file import map (Pass~1) then collects cross-file call edges from function bodies (Pass~2), using AST traversal for Python and regex for JavaScript/TypeScript. Each edge records two semantic flags: \texttt{return\_value\_used} (false for bare-expression / fire-and-forget calls) and \texttt{is\_awaited}. Edges are aggregated per HTTP entry point into a compact dependency map, reducing the Architect's search from O($n^2$) to O(1) per endpoint. Figure~\ref{fig:static-analysis} illustrates this process on a benchmark Flask application: a route handler delegating to a cross-file helper is analysed to produce the \texttt{entry\_point\_dependencies} record in \texttt{analysis\_report.json}.

\textbf{DynamoDB Schema Locator.} For applications tagged with \texttt{DynamoDB},
we identify schema definition files via a priority heuristic: initialization scripts
(e.g., \texttt{init\_db.py}, \texttt{create\_tables.js}) $\succ$ database
configuration modules (\texttt{db.py}, \texttt{models.py}) $\succ$ general
business logic. The top-3 candidate files are recorded in the report, enabling the
SAM Engineer to read exact table schemas rather than inferring them from handler code.

\subsection{Agent Tools and Domain Knowledge}
\label{sec:tools-knowledge}

Each agent is equipped with a role-specific tool set and a curated subset of the
domain knowledge base (Table~\ref{tab:agent-tools}). Tools provide structured, validated interfaces to the file system and external
validators, enabling agents to navigate and operate over the entire monolith
codebase, while also catching syntactically
invalid artifacts before they propagate downstream.
Domain knowledge supplements LLM training with up-to-date, task-specific guidance,
ensuring generated code and templates follow documented AWS best practices rather than
relying solely on the backbone LLM's training-data coverage of rapidly evolving
cloud APIs.

\textbf{File Tools.} We implement three file-interaction tools as a group:
\begin{itemize}[leftmargin=*,nosep]
    \item \textbf{ReadFileTool}: reads files with optional line-range selection;
    returns a truncation warning for large files ($>$500 lines) to prevent context
    overflow.
    \item \textbf{WriteFileTool}: writes files with built-in syntax validation
    (Python AST, JSON, YAML); a JSON merge-key mode allows agents to construct
    large structured documents section-by-section without exceeding context limits.
    \item \textbf{FileListTool}: lists directory contents with optional recursion,
    enabling agents to verify that all expected artifacts have been generated.
\end{itemize}

\textbf{CodeRAGTool.} Provides semantic code retrieval over the monolith source.
Offline, we build a RAG~\cite{lewis2020rag} vector index over source code chunks using CodeBERT~\cite{codebert2020}
(\texttt{microsoft/codebert-base}) embeddings and LlamaIndex, consuming the
\texttt{symbol\_table.json} produced by static analysis as the chunking reference.
At runtime, the tool returns top-$k$ raw code snippets with relevance scores and
file/line metadata. To avoid \emph{double inference}, the tool operates in
\emph{retriever-only} mode that returns raw snippets without LLM synthesis and
leaves all semantic interpretation to the agent's own inference call.

\textbf{SAMValidateTool.} Wraps \texttt{cfn-lint}~\cite{cfn-lint} (AWS CloudFormation Linter) and returns structured feedback distinguishing fatal errors from warnings, enabling an in-task validation--fix loop without leaving the inference context.

\textbf{Domain Knowledge Base.} Four reference documents compiled from the AWS SAM and Lambda Developer Guides~\cite{aws-sam-guide}: (1)~\emph{Basic Serverless Architecture} (resource types, API Gateway, Cognito); (2)~\emph{Async Patterns} (SQS~\cite{aws-sqs}/EventBridge~\cite{aws-eventbridge} event-source mapping); (3)~\emph{Lambda Coding Reference} (handler signatures, Layer packaging, env-var conventions); (4)~\emph{SAM Template Reference} (valid YAML properties, documented anti-patterns). Each document is chunked and indexed per agent following \emph{least-exposure}: each agent receives only the knowledge relevant to its task (Table~\ref{tab:agent-tools}).

\begin{table}[htbp]
\centering
\caption{Tool and Knowledge Base Assignment per Agent}
\label{tab:agent-tools}
\small
\begin{tabularx}{\columnwidth}{@{}lXX@{}}
\toprule
\textbf{Agent} & \textbf{Tools} & \textbf{Knowledge Bases} \\
\midrule
Architect      & ReadFile, WriteFile
               & Basic Architecture, Async Patterns \\
Code Developer & ReadFile, CodeRAG, WriteFile
               & Lambda Coding, Async Patterns \\
SAM Engineer   & ReadFile, WriteFile, FileList, SAMValidate
               & SAM Template Ref. \\
Validator      & ReadFile, WriteFile, FileList, SAMValidate
               & SAM Ref., Lambda Coding, Async \\
\bottomrule
\end{tabularx}
\end{table}

\subsection{Architect}
\label{sec:architect}

The Architect consumes \texttt{analysis\_report.json} and produces
\texttt{blueprint.json}, the single source of truth for all downstream agents.
The blueprint specifies Lambda function boundaries, inter-Lambda communication
patterns, shared infrastructure resources, and the authentication strategy.
Its top-level structure is:

\begin{itemize}[leftmargin=*,nosep]
  \item \texttt{lambda\_functions}: one entry per business endpoint, each recording
    trigger type, runtime, source files, auth requirement, and async publish targets
  \item \texttt{dynamodb\_tables}, \texttt{s3\_buckets}, \texttt{cognito},
    \texttt{api\_gateway}: shared infrastructure specifications
  \item \texttt{sqs\_queues}, \texttt{eventbridge\_rules},
    \texttt{lambda\_invoke\_permissions}: inter-Lambda communication resources
  \item \texttt{dropped\_functions}: auth endpoints and infrastructure scripts
    replaced by managed services
\end{itemize}

\textbf{Static-Analysis-Driven Endpoint Planning.}
The Architect reads the \texttt{entry\_points} array from the report to enumerate
all HTTP routes, and consults \texttt{file\_tags} to detect auth-related files.
Endpoints bearing standard or role-based auth decorators (e.g., \texttt{@login\_required},
\texttt{@warehouse\_required}, etc.) are classified as \texttt{auth="required"};
auth endpoints (\texttt{/register}, \texttt{/login}, \texttt{/logout}) are
\emph{dropped entirely} and delegated to AWS Cognito, following the
``infrastructure-over-code'' principle. The remaining business endpoints are
mapped one-to-one to Lambda functions (one Lambda per HTTP method+path), providing
least-privilege IAM scoping and independent scaling per endpoint.

\textbf{Source-Code-Guided Communication Pattern Selection.}
Assigning an inter-Lambda communication mechanism is a two-phase process.
In \emph{Phase~1}, the Architect uses static analysis hints from
\texttt{entry\_point\_dependencies}: \texttt{return\_value\_used=true} signals
that the caller requires the callee's result (favouring synchronous invoke),
while \texttt{return\_value\_used=false} combined with \texttt{is\_awaited=false} signals
a fire-and-forget side-effect (favouring async). However, \texttt{entry\_point\_dependencies}
captures only level-1 calls (route handler $\to$ service); cross-service calls
(service~A $\to$ service~B) are invisible at this level.
In \emph{Phase~2}, the agent therefore autonomously reads the relevant business
service source files to trace deeper call chains and resolve ambiguous cases.
For each identified cross-domain relationship, the agent applies three rules:
(1)~\emph{Synchronous Lambda Invoke}, when the caller needs the callee's return
value to construct its HTTP response; (2)~\emph{SQS queue}, for fire-and-forget
side-effects with a single consumer; and (3)~\emph{EventBridge rule}, when a single
action fans out to two or more independent consumers across different service domains.

\subsection{Code Developer}
\label{sec:code-developer}

The Code Developer reads \texttt{blueprint.json} and generates one Lambda handler
per \texttt{lambda\_functions} entry. For each Lambda, the blueprint provides:
the monolith \texttt{source\_files} to read, the trigger type and runtime, the
auth requirement, async publish targets (\texttt{publishes\_to}), and required
environment variable names. The agent reads the designated source files first;
for implementation details not covered by \texttt{source\_files} (helper
utilities, shared libraries), it queries \texttt{CodeRAGTool} and follows up
with \texttt{ReadFileTool} for full context.

\textbf{Code Transformations.} Three key adaptations are applied:
(1)~\emph{Framework adaptation}: Flask/Express route handlers are rewritten to
the \texttt{lambda\_handler(event, context)} signature, extracting path parameters
from \texttt{event['pathParameters']} and bodies from \texttt{event['body']};
(2)~\emph{Identity model adaptation}: the monolith's integer \texttt{user\_id}
references are replaced with the Cognito Sub UUID obtained from
\path{event['requestContext']['authorizer']['claims']['sub']}, and the
\texttt{Users} DynamoDB table is removed, and credentials are delegated to Cognito
while any business profile fields are retained in a separate table;
(3)~\emph{State elimination}: global variables are replaced by environment
variables injected at deploy time.

\textbf{Shared Layer and Dependency Management.}
The agent creates a shared Lambda Layer when utilities are reused across $\geq$3 Lambdas. Layer content is placed in the runtime-specific path (\texttt{python/} or \texttt{nodejs/}) required by the AWS Lambda Developer Guide~\cite{aws-lambda-layers}, so Lambda mounts it under \texttt{/opt/}. Dependency files are generated only for non-empty filtered dependency sets (stripping pre-bundled runtime packages such as \texttt{boto3} for Python; helper SDK packages must be explicitly declared for \texttt{nodejs22.x}~\cite{aws-lambda-nodejs}). All Node.js handlers use CommonJS (\texttt{require}) since Lambda's \texttt{/opt/nodejs} path is resolved via \texttt{NODE\_PATH}, which ESM does not honour.

\textbf{Cross-Lambda Communication.}
For those Lambdas with \texttt{publishes\_to} or \texttt{lambda\_invoke\_permissions} in the blueprint, the agent injects the corresponding SDK call (e.g., \path{lambda.invoke}, \texttt{sqs.send\_message}, or \texttt{events.put\_events}). Async publishes are \texttt{await}ed before the HTTP response is returned, since Lambda freezes the environment immediately on handler return~\cite{aws-lambda-lifecycle}.

\subsection{SAM Engineer}
\label{sec:sam-engineer}

The SAM Engineer generates \texttt{template.yaml} from the pre-defined \texttt{blueprint.json} and the generated code via three sequential sub-tasks following a \emph{dependency-based topological layering} principle. Sub-task~1 establishes shared stateful resources first (DynamoDB tables, Cognito UserPool, Lambda Layers, API Gateway), creating the CloudFormation logical reference anchors (\texttt{!Ref}, \texttt{!GetAtt}) that downstream definitions resolve. Sub-task~2 processes Lambda function resources domain-by-domain, appending incrementally to ensure referential integrity. Sub-task~3 wires EventBridge rule targets, appends the Outputs section, and closes with a \texttt{SAMValidateTool} validation loop that iteratively corrects errors before finalisation. Template correctness is critical: a single misconfigured property blocks the entire deployment pipeline, requiring the deep AWS-specific knowledge encoded in the domain knowledge base (\S\ref{sec:tools-knowledge}).

\subsection{Consistency Validator}
\label{sec:validator}

The Consistency Validator performs cross-artifact verification between Lambda code, SAM template, and blueprint, automatically fixing mismatches to ensure deployment readiness. It executes 11 checks across five phases: \textbf{(A) Structural}: bidirectional directory/function coverage and \texttt{CodeUri} path validity; \textbf{(B) Code-to-Template Alignment}: handler name validity, environment variable completeness, IAM policy alignment with SDK calls, and shared-layer reference consistency; \textbf{(C) Blueprint Matching}: API routes or methods matching against \texttt{blueprint.lambda\_functions} and per-endpoint auth override correctness; \textbf{(D) Async Resources}: SQS queue existence, producer/consumer policy and env var pairs, EventBridge rule targets and invocation permissions; \textbf{(E) Runtime Compliance}: dependency file correctness (stripping Lambda built-ins) and Node.js CJS/ESM enforcement. After all checks, fixes are applied in a single batch and re-validated with \texttt{SAMValidateTool}; anti-loop safeguards prevent infinite cycles.

\section{Benchmark}
\label{sec:benchmark}

Real-world monolithic applications rarely
come with paired serverless counterparts, which makes rigorous evaluation difficult.
We therefore adopt a \emph{reverse-engineering approach}. Starting from AWS-maintained
reference serverless applications in \texttt{aws-samples}~\cite{aws-samples}, we examined
the top 270 repositories and applied inclusion criteria (SAM-based, HTTP API, no
AI/ML dependencies, no proprietary external services). The screening yielded 12
candidates, from which we selected 6 that cover diverse complexity levels and the
two dominant Lambda runtimes, Python and JavaScript~\cite{icsa25llms}.

For each serverless application, we constructed an equivalent monolith by
(1) consolidating Lambda handlers into Flask/Express, (2) retaining DynamoDB for
application data to isolate migration complexity from database concerns,
(3) replacing Cognito with JWT-based middleware, (4) converting SQS/EventBridge
async patterns to synchronous equivalents while preserving detectable async
signals for the pipeline to reconstruct, and (5) removing frontend code. We then
verified each constructed monolith through E2E testing against the original
serverless behavior. The original serverless applications are used only to build
the benchmark and identify target cloud-native properties; all migration methods
operate exclusively on monolith source code. Table~\ref{tab:benchmark} summarizes
the benchmark characteristics.

\begin{table}[htbp]
\centering
\caption{Benchmark Applications}
\label{tab:benchmark}
\small
\begin{tabular}{@{}llrrrrcc@{}}
\toprule
\textbf{ID} & \textbf{Application} & \textbf{Lang} & \textbf{LoC} & \textbf{EP$_\text{obs}$} & \textbf{Tbl} & \textbf{Auth} & \textbf{Async} \\
\midrule
B1 & Todo & JS & 601 & 6 & 2 & \cmark & \xmark \\
B2 & Shopping Cart & Py & 1,130 & 8 & 2 & \cmark & \xmark \\
B3 & Bookstore & JS & 1,349 & 11 & 4 & \cmark & \cmark \\
B4 & Coffee Shop & JS & 1,666 & 11 & 6 & \cmark & \cmark \\
B5 & Airline Booking & Py & 2,109 & 14 & 4 & \cmark & \cmark \\
B6 & E-commerce & Py & 3,623 & 26 & 6 & \cmark & \cmark \\
\midrule
\multicolumn{2}{l}{\textbf{Total}} & -- & 10,478 & 76 & 24 & 6/6 & 4/6 \\
\bottomrule
\end{tabular}
\smallskip
\begin{minipage}{\columnwidth}
\footnotesize
EP$_\text{obs}$ = Observable business endpoints (auth endpoints excluded).
Tbl = DynamoDB tables in constructed monolith.
Async = monolith contains detectable async patterns (\S\ref{sec:metrics}).
\end{minipage}
\end{table}

\section{Experimental Setup}
\label{sec:experiment}

\subsection{Research Questions}

We evaluate \textsc{Mono2Sls} through four RQs: \textbf{RQ1 (Deployability)} asks whether it can generate deployable applications without manual intervention; \textbf{RQ2 (Functional Correctness)} asks how faithfully it preserves monolith behavior relative to commercial AI assistants and LLM-based generation; \textbf{RQ3 (Ablation)} asks what the contribution of static-analysis-driven architecture planning is; and \textbf{RQ4 (Cloud-Native Design)} asks whether the generated applications adopt AWS-native patterns. For RQ1--2, we compare against Cursor, Claude Code, and an LLM-Baseline, with two backbone LLMs (Claude Sonnet 4.6, DeepSeek-V3.2) to confirm results are method-driven rather than model-driven.

\subsection{Baselines and Ablation}
\label{sec:baselines}

\subsubsection{Commercial Baselines}

\begin{figure}[t]
\begin{tcolorbox}[
  enhanced,
  colback=white,
  colframe=gray!55,
  boxrule=0.8pt,
  arc=2mm,
  title={\small\bfseries Baseline Migration Prompt},
  coltitle=white,
  colbacktitle=gray!65,
  toptitle=2pt, bottomtitle=2pt,
  top=0pt, bottom=0pt, left=0pt, right=0pt,
]
{\setlength{\fboxsep}{0pt}%
\noindent\colorbox{yellow!20}{\parbox{\linewidth}{%
  \vspace{5pt}%
  \hspace{8pt}{\small\textbf{You are a senior software engineer
  specializing in AWS Serverless architecture.}}\par\vspace{3pt}%
  \hspace{8pt}{\small Task: Migrate this monolithic web application
  to AWS SAM-based serverless architecture.}\par\vspace{3pt}%
  \hspace{8pt}{\small\textbf{Your task:}}\par\vspace{5pt}%
}}\par\nointerlineskip
\noindent\colorbox{white}{\parbox{\linewidth}{%
  \vspace{3pt}%
  \begin{enumerate}[leftmargin=2.2em,noitemsep,topsep=2pt,parsep=0pt]
  \small
  \item Analyze the application and determine which endpoints to
        migrate as Lambda functions.
  \item Each Lambda function handles one API endpoint.
  \item Generate a complete, deployable SAM project.
  \item Preserve all original API functionality.
  \item Use Node.js 22.x or Python 3.12 runtime.
  \item Do not generate any documentation files.
  \end{enumerate}%
  \vspace{1pt}%
}}\par\nointerlineskip
\noindent\colorbox{yellow!20}{\parbox{\linewidth}{%
  \vspace{5pt}%
  \hspace{8pt}{\small\textbf{The source application code is already
  in this workspace.}}\par\vspace{4pt}%
  \hspace{8pt}{\ttfamily\fontsize{7.5}{10}\selectfont%
    /output\par%
    \hspace{8pt}+{-}{-}~template.yaml\par%
    \hspace{8pt}\textbackslash{-}{-}~lambdas/\par%
    \hspace{20pt}\textbackslash{-}{-}~\textit{<function\_name>}/\par%
    \hspace{32pt}+{-}{-}~handler.py (or handler.js)\par%
    \hspace{32pt}\textbackslash{-}{-}~requirements.txt (or package.json)%
  }\par\vspace{5pt}%
}}}
\end{tcolorbox}
\caption{Baseline migration prompt provided verbatim to Cursor and Claude Code.}
\label{fig:baseline-prompt}
\end{figure}

We evaluate \textbf{Cursor} (Agent mode with Claude Sonnet 4.5 (Thinking)) and \textbf{Claude Code} (CLI with Claude Sonnet 4.6), following the methodology of RepoGenesis~\cite{repogenesis}: each baseline receives the complete monolith source code and a task-level migration prompt (Figure~\ref{fig:baseline-prompt}) but no pre-computed static analysis or domain knowledge. All tool-initiated permission requests and confirmation prompts were accepted without redirecting agent decisions.

\subsubsection{LLM-Based Baseline}
\label{sec:llm-baseline}

To isolate the effect of staged decomposition, we build an \emph{LLM-based baseline}
(\textbf{LLM-Baseline}) that handles the migration in a single generalist context.
It receives the same inputs as \textsc{Mono2Sls} (static analysis, domain knowledge,
tool access) but replaces four specialized stages with one generalist agent,
collapses seven tasks into two, removes the 11-point consistency validator, and
replaces detailed AWS-specific prompts with high-level instructions. Implemented
with an agent framework to handle large codebases, it is essentially an
\emph{end-to-end generation} setup without iterative refinement. We evaluate two
backbone LLMs, DeepSeek-V3.2 (\textbf{LLM-Baseline-DS}) and Claude Sonnet 4.6
(\textbf{LLM-Baseline-SN}), to keep the comparison controlled on model choice.

\subsubsection{Ablation: Without Static-Analysis-Driven Architecture Planning}
\label{sec:ablation-design}

For the ablation, we remove both the static analyzer and the Architect agent.
These components are tightly coupled because \texttt{analysis\_report.json} is
designed specifically to support blueprint generation. The ablation tests whether
pre-computed structural analysis plus explicit architectural planning adds value
beyond direct source-to-target transformation. Without them, the Code Developer
must identify endpoints from source code, the SAM Engineer must infer resources
from generated handlers, and the Consistency Validator loses its architectural
ground-truth reference.

\begin{table}[htbp]
\centering
\caption{Comparison of experimental conditions}
\label{tab:conditions}
\small
\begin{tabular}{@{}lp{1.5cm}p{1.55cm}p{1.55cm}@{}}
\toprule
\textbf{Dimension} & \textbf{M2S (Full)} & \textbf{LLM-Baseline} & \textbf{w/o SA\&Arch} \\
\midrule
Static Analysis & \cmark & \cmark & \xmark \\
Architect Agent & \cmark & {\small(single)} & \xmark \\
Blueprint & \cmark & \cmark & \xmark \\
Agents & 4 spec. & 1 gen. & 3 spec. \\
Tasks (SAM) & 7 (3) & 2 & 6 (2) \\
Validator & 11-point & Removed & Partial \\
Domain Knowledge & \cmark & \cmark & \cmark \\
Prompt Guidance & Detailed & High-level & Detailed \\
\bottomrule
\end{tabular}
\smallskip
\begin{minipage}{\columnwidth}
\footnotesize
spec.~= specialized; gen.~= generalist; Tasks~(SAM)~= total tasks (SAM sub-tasks).
\end{minipage}
\end{table}

\subsection{Evaluation Metrics}
\label{sec:metrics}

We define metrics across two dimensions: deployability and functional correctness.

\textbf{Deployability.} We measure deployability through two complementary metrics:

\emph{Validation Pass Rate (VPR)} measures the fraction of generated projects that pass static validation checks. We use \texttt{sam validate}, which verifies SAM template syntax and resource references, together with \texttt{cfn-lint}, which enforces AWS CloudFormation best practices and detects common misconfigurations.

\emph{Deployment Success Rate (DSR)} measures the fraction of projects that achieve functional deployment to AWS: (1) \texttt{sam build} completes without error, (2) \texttt{sam deploy} provisions all resources without CloudFormation rollback, and (3) deployed Lambdas are invocable without systematic failures(e.g., universal HTTP 502 from incorrect handler paths or package structures). We manually inspect invocation errors to distinguish systematic failures from isolated bugs.

\textbf{Functional Correctness.} We assess functional quality through two complementary metrics that measure structural and behavioral fidelity respectively.

\emph{API Coverage (F1)} measures the match between generated HTTP endpoints $G$ and the reference set $R$ of observable business endpoints from the ground-truth monolith ($|R|=76$). It is computed as the harmonic mean of precision (how many generated endpoints belong in $R$) and recall (how many reference endpoints are generated). Redundant auth endpoints (not in $R$) reduce precision; async consumer Lambdas (no HTTP surface) are excluded from both sets.

\emph{End-to-End Pass Rate (E2EPR)} measures the fraction of E2E tests passing against the deployed application, capturing behavioral correctness beyond structural compliance. Let $T_i$ denote the test suite for application $a_i$ and $P_i$ the subset of passing tests:
\begin{equation}
\text{E2EPR}_{\text{micro}} = \frac{\sum_i |P_i|}{\sum_i |T_i|}, \quad
\text{E2EPR}_{\text{macro}} = \frac{1}{n}\sum_i \frac{|P_i|}{|T_i|}
\end{equation}
We report on Core + Robustness tests (121 out of 145 total); test categories and their applicability to each method are detailed in \S\ref{sec:testsuite}. Unless otherwise stated, $T_i$ refers to Core + Robustness tests applicable to application $a_i$.

\subsection{Test Suite and Environment}
\label{sec:testsuite}

We design application-agnostic E2E test suites by extracting API contracts from each monolith (endpoint signatures, request/response schemas, and authentication requirements) and using GPT-5.4, separate from the pipeline LLMs, to generate 145 tests covering four categories: \emph{Core} (CRUD), \emph{Robustness} (edge cases), \emph{Auth} (JWT endpoints), and \emph{Async} (event-driven). All tests then undergo expert review to ensure correctness and coverage. To evaluate methods with different authentication mechanisms under a common interface, we implement an \emph{Auth Provider} abstraction (\texttt{CognitoProvider} for \textsc{Mono2Sls}; \texttt{CustomJwtProvider} for baselines), so the test logic remains auth-agnostic. The \textbf{Core+Robustness tests (121)} serve as the primary metric for all methods; Auth tests (18) apply only to commercial baselines; Async tests (6) apply only to methods that generate event-driven architectures (qualitative analysis in \S\ref{sec:rq4}).

\begin{table}[htbp]
\centering
\caption{Test Suite Distribution}
\label{tab:testsuite}
\small
\begin{tabular}{@{}lrrrrr@{}}
\toprule
\textbf{App} & \textbf{Core} & \textbf{Rob} & \textbf{Auth} & \textbf{Async} & \textbf{Total} \\
\midrule
B1: Todo & 13 & 3 & 0 & 0 & 16 \\
B2: Shopping Cart & 12 & 3 & 2 & 0 & 17 \\
B3: Bookstore & 12 & 4 & 4 & 1 & 21 \\
B4: Coffee & 19 & 2 & 3 & 1 & 25 \\
B5: Airline & 19 & 5 & 6 & 2 & 32 \\
B6: E-commerce & 24 & 5 & 3 & 2 & 34 \\
\midrule
\textbf{Total} & 99 & 22 & 18 & 6 & 145 \\
\bottomrule
\end{tabular}
\end{table}

\textbf{Environment.} Pipeline execution runs locally (Windows). Generated applications 
are deployed to AWS via \texttt{sam deploy} and tested against live infrastructure 
(API Gateway, Lambda, DynamoDB). Both backbone LLMs (DeepSeek-V3.2 and Claude Sonnet 4.6) 
were run at their default inference temperature; all experiments report single-run 
(pass@1) results per application.

\section{Results}
\label{sec:results}

We report results for each research question in turn, following the evaluation protocol and metrics defined in \S\ref{sec:experiment}.

\textbf{Evaluation Protocol.} Because the methods differ in deployment success, we
use different evaluation procedures. \textbf{Commercial baselines} receive
expert-assisted fixes for both validation and deployment failures, which isolates
the comparison to functional correctness and reflects their best-case usage
scenario. \textbf{LLM-Baseline and Ablation} receive fixes only for validation
failures (VPR=0) so they can be deployed for testing, but systematic deployment
failures (DSR=0) remain unfixed and are marked non-functional (E2EPR = N/A).
\textsc{\textbf{Mono2Sls}} is evaluated on raw output only, because the
self-verifying validator is intended to ensure deployment readiness. Under this
setup, commercial tools are assessed under their intended collaborative usage
model while the research variants are assessed on raw generation quality.

\subsection{RQ1: Deployability}
\label{sec:rq1}

\begin{table}[htbp]
\centering
\caption{Deployment Results (Raw vs.\ Evaluated)}
\label{tab:deployment}
\small
\begin{tabular}{@{}lcccc@{}}
\toprule
\textbf{Method} & \textbf{VPR(raw)} & \textbf{VPR(eval)} & \textbf{DSR(raw)} & \textbf{DSR(eval)} \\
\midrule
M2S-DeepSeek      & 6/6 & 6/6 & 6/6 & 6/6 \\
M2S-Sonnet        & 6/6 & 6/6 & 6/6 & 6/6 \\
LLM-Baseline-DS   & 6/6 & 6/6 & 3/6 & -- \\
LLM-Baseline-SN   & 6/6 & 6/6 & 3/6 & -- \\
Cursor            & 2/6 & \textit{6/6}$^{\dagger}$ & 4/6 & \textit{6/6}$^{\dagger}$ \\
Claude Code       & 3/6 & \textit{6/6}$^{\dagger}$ & 4/6 & \textit{6/6}$^{\dagger}$ \\
\bottomrule
\end{tabular}
\smallskip

\begin{minipage}{\columnwidth}
\footnotesize
raw = output without any fixes. eval = after applying evaluation-protocol fixes.
\textit{Italic} values differ from raw due to applied fixes.
LLM-Baseline DSR failures are not fixed per evaluation protocol (\S\ref{sec:metrics}).
$^{\dagger}$Commercial baselines receive expert-assisted fixes for both validation 
and deployment failures.
\end{minipage}
\end{table}

Table~\ref{tab:deployment} reports deployment results. \textsc{Mono2Sls} reaches
100\% VPR and DSR for both backbone models, without manual fixes.

Commercial baselines produce invalid templates on 2--4 of the 6 applications.
Common issues include reserved environment variables, invalid CORS settings, and
unsupported \texttt{Globals} combinations. We do not observe these errors in
\textsc{Mono2Sls} outputs, likely because generation is constrained by the SAM
knowledge base. After expert-assisted validation fixes, both Cursor and Claude
Code still fail to deploy 2 applications because of incorrect Lambda Layer
directory structures. \textsc{Mono2Sls} avoids these failures through agent
specialization and the Consistency Validator's cross-artifact checks.

Both LLM-Baseline variants achieve 6/6 VPR but only 3/6 DSR. Their failures arise
from Cognito schema mis-declarations and API definition errors that are
syntactically valid yet semantically wrong. cfn-lint does not catch these faults,
but Mono2Sls's Consistency Validator does through cross-artifact comparison with
the blueprint and handler code. The same E-commerce identity configuration error
appears under both backbone LLMs, which points to a limitation of single-context
generation rather than a model-specific weakness: without separate phases for
architecture design, code generation, and template validation, the system cannot
reliably maintain structural coherence across all generated artifacts.
LLM-Baseline DSR values are reported as-is per the evaluation protocol
(\S\ref{sec:metrics}); no deployment fixes are applied.

\textsc{Mono2Sls} is the only method that reaches 100\% VPR and DSR without any
manual intervention. The commercial baselines require expert-assisted fixes, and
the LLM-Baseline variants still fail to deploy 3 of 6 applications despite
producing structurally valid templates.

\subsection{RQ2: Functional Correctness}
\label{sec:rq2}

\subsubsection{API Coverage}

\begin{table}[htbp]
\centering
\caption{API Coverage F1 Scores}
\label{tab:api-coverage}
\small
\begin{tabular}{@{}lcccc@{}}
\toprule
\textbf{Method} & \textbf{Micro-F1} & \textbf{Macro-F1} & \textbf{Anti-Pat.} & \textbf{Consumer} \\
\midrule
M2S-DeepSeek      & 1.000  & 1.000  & 0  & 11 \\
M2S-Sonnet        & 0.987  & 0.986  & 1  & 2  \\
LLM-Baseline-DS   & 0.9935 & 0.9928 & 1  & 0  \\
LLM-Baseline-SN   & 0.974  & 0.969  & 2  & 0  \\
Cursor            & 0.884  & 0.870  & 19 & 0  \\
Claude Code       & 0.889  & 0.879  & 19 & 0  \\
\bottomrule
\end{tabular}
\smallskip

\begin{minipage}{\columnwidth}
\footnotesize
Anti-Pat. = Redundant auth endpoints. Consumer = Async consumer Lambdas (SQS/EventBridge-triggered, 
correctly excluded from API surface).
\end{minipage}
\end{table}

Table~\ref{tab:api-coverage} presents API Coverage F1 scores. Mono2Sls achieves 
near-perfect F1 (1.000 for M2S-DeepSeek, 0.987 for M2S-Sonnet), compared to 
LLM-Baseline variants (DS: 0.994, SN: 0.974) and commercial baselines (0.884--0.889).

Much of the gap comes from authentication design. By delegating authentication
to AWS Cognito, Mono2Sls does not generate redundant \texttt{/register},
\texttt{/login}, and \texttt{/auth/me} endpoints. Commercial baselines generate
19 such extra APIs across the 6 applications, while LLM-Baseline-SN generates 2
and LLM-Baseline-DS generates 1; the smaller counts for the LLM baselines
reflect the same domain knowledge that instructs the model to leave
authentication to Cognito.

The LLM-Baseline variants still achieve high API F1, which indicates that static
analysis input and domain knowledge are often enough for a single-context model
to recover the API surface correctly. Yet recovering the interface is not enough
to ensure correct behaviour, as the next subsection shows.

\subsubsection{End-to-End Correctness}

\begin{table}[htbp]
\centering
\caption{End-to-End Pass Rate (\%) on Core + Robustness Tests}
\label{tab:e2e}
\resizebox{\columnwidth}{!}{%
\begin{tabular}{@{}lcccccc|cc@{}}
\toprule
\textbf{Method} & \textbf{B1} & \textbf{B2} & \textbf{B3} & \textbf{B4} & \textbf{B5} & \textbf{B6} & \textbf{Micro} & \textbf{Macro} \\
\midrule
M2S-DeepSeek      & 100  & 33.3 & 93.8 & 71.4 & 50.0 & 51.7 & 64.5 & 66.7 \\
M2S-Sonnet        & 100  & 33.3 & 93.8 & 76.2 & 50.0 & 55.2 & \textbf{66.1} & \textbf{68.1} \\
LLM-Baseline-DS   & N/A  & 33.3 & 31.3 & N/A  & 37.5 & N/A  & 34.5$^*$ & 34.0$^*$ \\
LLM-Baseline-SN   & 37.5 & N/A  & N/A  & 23.8 & 41.7 & N/A  & 34.4$^*$ & 34.3$^*$ \\
Cursor            & 100  & 26.7 & 31.2 & 100  & 45.8 & 27.6 & 53.7 & 55.2 \\
Claude Code       & 100  & 33.3 & 31.2 & 100  & 87.5 & 20.7 & 61.2 & 62.1 \\
\bottomrule
\end{tabular}%
}
\smallskip

\begin{minipage}{\columnwidth}
\footnotesize
N/A = DSR=0 (LLM-Baseline, unfixed per protocol). $^*$Computed over 3 DSR-passing apps only (SN: B1, B4, B5; DS: B2, B3, B5).
\end{minipage}
\end{table}

Table~\ref{tab:e2e} compares end-to-end functional correctness. M2S-Sonnet
obtains the highest micro-E2EPR at 66.1\%, ahead of Claude Code (61.2\%) and
Cursor (53.7\%). The margin is notable because the commercial baselines were
allowed expert-assisted deployment fixes, whereas Mono2Sls was evaluated on raw
output. The remaining gap therefore reflects generated code quality rather than
deployment scaffolding.

\textbf{LLM-Baseline Performance.} Despite high API F1, both LLM-Baseline variants
achieve only about 34\% micro-E2EPR (34.5\% for LLM-Baseline-DS and 34.4\% for
LLM-Baseline-SN), computed over the three applications each variant deploys
successfully. The contrast exposes a \emph{structural-to-functional gap}: a
single-context LLM can often identify \emph{what} APIs to generate while still
failing to implement the underlying business logic, database access patterns, and
cross-function coordination correctly. The F1-to-E2EPR gap reaches about 65pp for
LLM-Baseline-DS (99.4\% F1 vs.\ 34.5\% micro-E2EPR) and about 63pp for
LLM-Baseline-SN (97.4\% vs.\ 34.4\%), compared with about 33pp for M2S-Sonnet
(98.7\% vs.\ 66.1\%). The smaller gap reflects the implementation benefit of
staged decomposition and specialized prompting.

\textbf{Backbone LLM Comparison.} M2S-Sonnet and M2S-DeepSeek produce similar
results: E2EPR differs by 1.6pp (66.1\% vs.\ 64.5\%), and both reach 6/6 VPR
and DSR. In our setting, pipeline design appears to matter more than the choice
of backbone model. The same pattern appears in the LLM baselines, where SN and
DS reach nearly identical E2EPR (34.4\% vs.\ 34.5\%). The roughly 32pp gap
between Mono2Sls and the LLM baselines is therefore more plausibly explained by
pipeline structure than by model strength.

\subsection{RQ3: Contribution of Static-Analysis-Driven Architecture Planning}
\label{sec:rq3}

We compare the ablation variant (w/o SA\&Arch) against M2S-Sonnet (the same backbone 
LLM used in the ablation) across deployability, API coverage, and end-to-end correctness.

\begin{table}[htbp]
\centering
\caption{Ablation Results: M2S-Sonnet vs.\ w/o SA\&Arch}
\label{tab:ablation-results}
\small
\begin{tabular}{@{}lcccc@{}}
\toprule
\textbf{App} & \textbf{DSR} & \textbf{F1} & \textbf{E2EPR} & \textbf{Failure Mode} \\
\midrule
B1: Todo          & \cmark & 0.833 & 56.3\% & None \\
B2: Shopping Cart & \cmark & 1.000 & 20.0\% & None \\
B3: Bookstore     & \xmark & 1.000 & N/A    & Env var not injected \\
B4: Coffee        & \xmark & 0.957 & N/A    & Path variable conflict \\
B5: Airline       & \xmark & 0.714 & N/A    & Layer double-nesting \\
B6: E-commerce    & \cmark & 0.962 & 37.9\% & None \\
\midrule
\textbf{w/o SA\&Arch} & 3/6 & 0.915 & 38.3\%$^*$  & \\
\textbf{M2S-Sonnet}   & 6/6 & 0.987 & 66.1\%$^{\dagger}$ & \\
\bottomrule
\end{tabular}
\smallskip

\begin{minipage}{\columnwidth}
\footnotesize
VPR = 6/6 for both variants. M2S-Sonnet F1 and overall E2EPR from Tables~\ref{tab:api-coverage}--\ref{tab:e2e}.\\
$^*$micro-E2EPR over 3 DSR-passing applications (B1, B2, B6).\\
$^{\dagger}$Overall micro-E2EPR (all 6 apps); on the same 3-app subset (B1, B2, B6), M2S-Sonnet achieves 61.7\%.
\end{minipage}
\end{table}

Table~\ref{tab:ablation-results} shows a clear drop in performance without static
analysis and architectural planning.

\textbf{Deployability (DSR: 6/6$\to$3/6).} The root cause is the absence of 
\texttt{blueprint.json} as a shared contract: without a centrally authored 
endpoint inventory and resource specification, each agent must independently infer 
structure from source code, leading to cross-artifact inconsistencies that survive 
template validation but break at deployment. Concretely, all three DSR failures 
involve handler--template mismatches that \texttt{cfn-lint} cannot detect: missing 
environment variable injections, inconsistent API path parameters across sibling 
routes, and incorrect Lambda Layer directory structures. These are precisely the 
failure modes the Architect and Consistency Validator are designed to prevent.

\textbf{API Coverage (F1: 0.987$\to$0.915).} The F1 drop reflects
\emph{REST-shape drift}, not only missing endpoints. Applications with flat API
structures (B2, B3) retain perfect F1, whereas those with more complex routing
(B1, B5) regress the most. The drift appears as resource hierarchy changes
(e.g., B5: \texttt{/customers/\{id\}/bookings} $\to$
\texttt{/bookings/customer/\{id\}}), HTTP method reassignment (B5:
\texttt{POST} $\to$ \texttt{PUT} for state transitions, leading to HTTP 405
errors for conforming clients), and path compression in the multi-domain B6
application. Static analysis therefore does more than point to relevant files; it
constrains the space of plausible API designs.

\textbf{End-to-End Correctness.} On the 3 applications deployable by both
variants (B1, B2, B6), M2S-Sonnet reaches 61.7\% micro-E2EPR (computed from
Table~\ref{tab:e2e}: $(16{+}5{+}16)/(16{+}15{+}29)$), compared with 38.3\% for
w/o SA\&Arch, a 23.4pp gap on the same application set. Without
blueprint-defined Lambda boundaries and dependency relationships, the Code
Developer must read the monolith, infer endpoint semantics, and generate
handlers at the same time. That broader context hurts implementation accuracy
even when deployment still succeeds.

Overall, removing static-analysis-driven architecture planning degrades all
three quality dimensions: DSR (6/6$\to$3/6), API Coverage F1
(0.987$\to$0.915), and micro-E2EPR on the comparable deployable subset
(61.7\%$\to$38.3\%). The ablation therefore supports the claim that
pre-computed structural analysis plus explicit architectural planning adds value
beyond direct source-to-target generation.

\subsection{RQ4: Cloud-Native Design Adoption}
\label{sec:rq4}

\begin{figure*}[!t]
    \centering
    \includegraphics[width=0.9\textwidth]{figure3}
    \caption{Case Study: architecture contrast on the Coffee Shop application (RQ4).
    (a)~Authentication: Mono2Sls routes requests through API Gateway with a Cognito
    Authorizer, eliminating custom auth logic from Lambda handlers; commercial baselines
    replicate the monolith's JWT middleware pattern.
    (b)~Asynchronous communication: Mono2Sls builds a complete SQS event-driven chain
    with decoupled producer and consumer Lambdas; commercial baselines produce
    synchronous direct writes.}
    \Description{Four-panel architecture diagram comparing Mono2Sls and commercial baselines on the Coffee Shop application. Top row: authentication architecture. Mono2Sls uses API Gateway with Cognito Authorizer (left), while baselines retain custom JWT middleware in Lambda handlers (right). Bottom row: async communication. Mono2Sls generates an SQS-based event-driven pipeline with producer and consumer Lambdas (left), while baselines use synchronous direct writes to DynamoDB (right).}
    \label{fig:rq4-casestudy}
\end{figure*}

Beyond functional correctness, we also examine whether the generated
applications adopt AWS-native architectural patterns in two areas.
\textbf{(1) Authentication}: AWS Cognito handles user registration, token
issuance, and token validation as a managed service; API Gateway authorizers can
consume Cognito tokens directly, removing authentication logic from Lambda
handlers. \textbf{(2) Asynchronous communication}: AWS SQS and EventBridge let
producers and consumers communicate through queues or event buses, which better
matches the stateless execution model of serverless systems.

\subsubsection{Authentication Architecture}

\textsc{Mono2Sls} uses AWS-native authentication in all 6 applications for both
backbone models. API Gateway Cognito authorizers~\cite{aws-cognito} protect the
routes, handlers read validated claims from the authorizer context, and no
custom authentication logic remains in the generated code. In contrast, the
commercial baselines retain the monolith's JWT-based design in all 6
applications for both Cursor and Claude Code, including application-managed
token validation, explicit \texttt{/register}/\texttt{/login} endpoints, and
developer-maintained middleware. Figure~\ref{fig:rq4-casestudy}~(a) shows this
contrast on the Coffee Shop application.

\subsubsection{Asynchronous Communication}

Across the 4 applications with detectable event-driven signals in the monolith
source (B3--B6), \textsc{Mono2Sls} often reconstructs full event-driven chains.
M2S-DeepSeek does so in all 4 applications; M2S-Sonnet does so in 2 and
produces partial async structure in the other 2 (producer wired, but not all
consumers generated). In both variants, the Architect records the communication
pattern in \texttt{blueprint.json}, and the downstream agents implement it in
code and infrastructure. By contrast, the commercial baselines generate purely
synchronous architectures across all 4 applications: neither Cursor nor Claude
Code attempts event-driven integration despite clear async signals in the
source. Figure~\ref{fig:rq4-casestudy}~(b) shows the difference on the Coffee
Shop application.

\subsubsection{LLM-Baseline: Hybrid Architecture and Responsibility Confusion}

LLM-Baseline variants use AWS domain knowledge, but they often produce
\emph{hybrid} designs that mix cloud-native and monolith-style components
within one application.

Qualitative inspection suggests that this pattern comes from blurred
architectural responsibilities. First, several applications provision Cognito
User Pools while still keeping redundant \texttt{Users} DynamoDB tables, which
shows that the model knows \emph{how} to add Cognito but not when it should
replace custom identity storage. Second, some applications declare Cognito
resources but omit or misconfigure API Gateway authorizers, leaving part of the
authentication logic inside handlers. Third, some async-capable applications
fall back to synchronous implementations: queues appear in the template, but no
consumer Lambdas are generated. LLM-Baseline-DS even combines correct async
adoption with incorrect authentication redesign in the same application. This
mixed outcome is consistent with a single-context setup that has to resolve
several architectural concerns at once. Mono2Sls avoids this by assigning each
concern to a narrower agent context.

\section{Discussion}
\label{sec:discussion}

\textsc{Mono2Sls} is intended for teams maintaining monolithic web backends under
scalability and operational cost pressures. Given a monolith repository, the
pipeline can produce a deployable SAM application, including Lambda handlers,
infrastructure template, and shared layers, without requiring cloud expertise
from the developer. M2S-Sonnet required 58--106~minutes per benchmark
application (601--3,623 LoC), with execution time dominated by the Code
Developer's iterative source transformation and scaling with application size
and backbone LLM choice. By automating architectural decomposition,
authentication transformation, and IaC generation, \textsc{Mono2Sls} reduces
what practitioners often describe as weeks of manual refactoring effort to
hours of automated processing per application.

\section{Limitations and Threats to Validity}
\label{sec:limitations}

\textbf{Benchmark Scope.} Our benchmark contains 6 applications derived from AWS
reference architectures. Although it covers both major serverless languages and
a range of sizes (601--3,623 LoC, 6--26 observable business endpoints), it does
not cover every monolith pattern, especially legacy enterprise systems with
heavy ORM usage or multiple databases. Future evaluation should include larger
industrial applications.

\textbf{AWS-Specific.} Mono2Sls currently targets AWS SAM. The core ideas in the
pipeline, namely static analysis, staged decomposition, and domain
knowledge, should transfer to other platforms, but support for Azure
Functions or Google Cloud Run would require platform-specific knowledge bases
and generators.

\textbf{Async Correctness.} While Mono2Sls achieves 100\% structural correctness
for async design (RQ4), only 67\% of async end-to-end tests pass for
M2S-Sonnet. We observe two main failure categories:
(1)~\emph{upstream confounding}, where earlier operations fail before the async
workflow is reached (e.g., payment integration in Airline Booking), and
(2)~\emph{eventual consistency bugs}, where producers and consumers are wired
correctly but still contain logic errors, such as incorrect event payload
schemas or missing idempotency keys. The first category reflects failures
outside the async mechanism itself; the second points to limits in current LLM
reasoning about distributed semantics.

\textbf{Internal Validity.} LLM inference is non-deterministic, and our pass@1
evaluation does not measure run-to-run variance. Single-run evaluation is
standard in repository-level code generation~\cite{swebench2024}, but it remains
a limitation. We expect variance to be smaller here than in open-ended
generation because the static-analysis inputs are deterministic and each agent
follows explicit task instructions with concrete AWS constraints. Still, only
repeated runs can confirm that expectation. Multi-run evaluation with
statistical testing is left to future work.

\section{Conclusion}
\label{sec:conclusion}

\textsc{Mono2Sls} is a static-analysis-driven multi-stage pipeline for automated
monolith-to-serverless migration. Rather than asking a single model to solve the
entire migration problem at once, Mono2Sls separates the work into four
sequential stages handled by specialized tool-using LLM agents and grounds their
decisions in static-analysis outputs and a curated AWS knowledge base. The
staged organization helps keep architectural
decomposition, code transformation, and infrastructure generation aligned more
reliably than single-context generation.

Across 6 benchmark applications (10K+ LoC, 76 business endpoints),
\textsc{Mono2Sls} achieves 100\% deployment success without manual intervention,
66.1\% micro-average end-to-end correctness, 98.7\% API-coverage F1, and
stronger adoption of AWS-native authentication and asynchronous patterns than
the baselines. Commercial baselines trail these results even after
expert-assisted deployment fixes, and the ablation study shows that
static-analysis-driven architecture planning adds 23.4 percentage points to
end-to-end correctness on the comparable deployable subset.

The main limitations are benchmark scale and AWS specificity. The current
pipeline also does not fully address stateful HTTP session patterns or
eventual-consistency edge cases in async workflows. Extending the approach to
other serverless platforms and improving LLM reasoning about distributed event
semantics are natural next steps.

\section*{Data Availability}
\label{sec:data-availability}
Our pipeline code, benchmarks, and test suites are available at: \url{https://doi.org/10.5281/zenodo.19230004}.

\begin{acks}
Parts of this paper were written and revised with the assistance of AI language models (including Claude by Anthropic). All experimental results, technical claims, figures, and tables are the authors' own work; AI tools were used solely for language editing, text organization, and literature search assistance.
\end{acks}

\newpage
\bibliographystyle{ACM-Reference-Format}
\bibliography{references}

\end{document}